\let\oldvec\vec % needed to suppress amsmath warning "cannot redefine \vec"
\documentclass[runningheads,a4paper]{llncs}
\let\vec\oldvec
\usepackage[utf8]{inputenc}
\usepackage[T1]{fontenc}
\usepackage{amsmath,amstext,amsopn,amssymb}
\usepackage{dsfont}
\usepackage{multirow}
\usepackage{color}
\usepackage{xifthen}
\usepackage{graphicx}
\usepackage{enumerate}
\usepackage{longtable}
\usepackage{mathrsfs}
\usepackage{bm} % for boldface math symbols
\usepackage{array}

\usepackage{float}

\newcolumntype{L}[1]{>{\raggedright\let\newline\\\arraybackslash\hspace{0pt}}m{#1}}
\newcolumntype{C}[1]{>{\centering\let\newline\\\arraybackslash\hspace{0pt}}m{#1}}
\newcolumntype{R}[1]{>{\raggedleft\let\newline\\\arraybackslash\hspace{0pt}}m{#1}}

\usepackage{hyperref}

\renewcommand{\vec}[1]{\bm{#1}}

\makeatletter
\newcommand\xleftrightarrow[2][]{%
	\ext@arrow 9999{\longleftrightarrowfill@}{#1}{#2}}
\newcommand\longleftrightarrowfill@{%
	\arrowfill@\leftarrow\relbar\rightarrow}
\makeatother

\usepackage{tikz}
\usetikzlibrary{positioning,arrows,shapes,backgrounds,calc,patterns}

\DeclareMathOperator{\HSR}{\mathsf{HSR}}
\DeclareMathOperator{\HSF}{\mathsf{HSF}}
\DeclareMathOperator{\HSFHSP}{\mathsf{HSPHSF}}

\DeclareMathOperator{\hsf}{\mathsf{hsf}}
\DeclareMathOperator{\hsfdimer}{\mathsf{hsf_2}}
\DeclareMathOperator{\hsftrimer}{\mathsf{hsf_3}}
\DeclareMathOperator{\hsfhse}{\mathsf{hsf_3:hse}}
\DeclareMathOperator{\hsp}{\mathsf{hsp}}
\DeclareMathOperator{\hsphsf}{\mathsf{hsp:hsf}}
\DeclareMathOperator{\hse}{\mathsf{hse}}
\DeclareMathOperator{\prot}{\mathsf{prot}}
\DeclareMathOperator{\mfp}{\mathsf{mfp}}
\DeclareMathOperator{\hspmfp}{\mathsf{hsp:mfp}}

\DeclareMathOperator{\rhsf}{\mathsf{rhsf}}

\DeclareMathOperator{\rhsfe}{\mathsf{rhsf(e)}}

\DeclareMathOperator{\rhsfdimer}{\mathsf{rhsf2}}
\DeclareMathOperator{\rhsfdimerf}{\mathsf{rhsf2(f)}}
\DeclareMathOperator{\rhsphsf}{\mathsf{rhsp:hsf}}

\DeclareMathOperator{\rhsfZ}{\mathsf{rhsf(HSF\_0)}}
\DeclareMathOperator{\rhsfO}{\mathsf{rhsf(HSF\_1)}}

\DeclareMathOperator{\rhsfdimertwo}{\mathsf{rhsf2(HSF2\_1)}}
\DeclareMathOperator{\HSFZ}{\mathsf{HSF\_0}}
\DeclareMathOperator{\HSFO}{\mathsf{HSF\_1}}
\DeclareMathOperator{\HSFTZ}{\mathsf{HSF2\_0}}
\DeclareMathOperator{\HSFTO}{\mathsf{HSF2\_1}}
\DeclareMathOperator{\HSFTT}{\mathsf{HSF2\_2}}
\DeclareMathOperator{\HSPFSFZ}{\mathsf{HSPHSF\_0}}
\DeclareMathOperator{\HSPFSFO}{\mathsf{HSPHSF\_1}}
\DeclareMathOperator{\RHSFdimer}{\mathsf{HSF2}}

\DeclareMathOperator{\rhsfzero}{\mathsf{\rhsf}^{(0)}}
\DeclareMathOperator{\rhsfone}{\mathsf{\rhsf}^{(1)}}

\DeclareMathOperator{\rhsfdimnac}{\mathsf{\rhsf_2^{(0)}}}
\DeclareMathOperator{\rhsfdimacone}{\mathsf{\rhsf_2^{(1)}}}
\DeclareMathOperator{\rhsfdimactwo}{\mathsf{\rhsf_2^{(2)}}}

\DeclareMathOperator{\hsprhsf}{\mathsf{\hsp:\rhsf}}
\DeclareMathOperator{\hsprhsfnac}{\mathsf{{\hsprhsf}^{(0)}}}
\DeclareMathOperator{\hsprhsfone}{\mathsf{{\hsprhsf}^{(1)}}}

\DeclareMathOperator{\EGFEGFRdim}{\mathsf{EGFEGFRdim}}
\DeclareMathOperator{\EGFEGFRrdim}{\mathsf{EGFEGFRrdim}}
\DeclareMathOperator{\EGFEGFRdimh}{\mathsf{EGFEGFRdim(h)}}

\DeclareMathOperator{\EGFErbBonedim}{\mathsf{EGFErbB1dim}}
\DeclareMathOperator{\EGFErbBtwodim}{\mathsf{EGFErbB2dim}}
\DeclareMathOperator{\EGFErbBthrdim}{\mathsf{EGFErbB3dim}}
\DeclareMathOperator{\EGFErbBfourdim}{\mathsf{EGFErbB4dim}}
\DeclareMathOperator{\HRGErbBonedim}{\mathsf{HRGErbB1dim}}
\DeclareMathOperator{\HRGErbBtwodim}{\mathsf{HRGErbB2dim}}
\DeclareMathOperator{\HRGErbBthrdim}{\mathsf{HRGErbB3dim}}
\DeclareMathOperator{\HRGErbBfourdim}{\mathsf{HRGErbB4dim}}

\DeclareMathOperator{\EGFEGFRxdim}{\mathsf{EGFEGFRxdim}}
\DeclareMathOperator{\EGFEGFRxrdim}{\mathsf{EGFEGFRxrdim}}
\DeclareMathOperator{\EGFEGFRxdimi}{\mathsf{EGFEGFRxdim(i)}}
\DeclareMathOperator{\EGFErbBonexdim}{\mathsf{EGFErbB1xdim}}
\DeclareMathOperator{\EGFErbBtwoxdim}{\mathsf{EGFErbB2xdim}}
\DeclareMathOperator{\EGFErbBthrxdim}{\mathsf{EGFErbB3xdim}}
\DeclareMathOperator{\EGFErbBfourxdim}{\mathsf{EGFErbB4xdim}}
\DeclareMathOperator{\HRGErbBonexdim}{\mathsf{HRGErbB1xdim}}
\DeclareMathOperator{\HRGErbBtwoxdim}{\mathsf{HRGErbB2xdim}}
\DeclareMathOperator{\HRGErbBthrxdim}{\mathsf{HRGErbB3xdim}}
\DeclareMathOperator{\HRGErbBfourxdim}{\mathsf{HRGErbB4xdim}}

\DeclareMathOperator{\ErbB}{\mathsf{ErbB}}
\DeclareMathOperator{\ErbBone}{\mathsf{ErbB1}}
\DeclareMathOperator{\ErbBtwo}{\mathsf{ErbB2}}
\DeclareMathOperator{\ErbBthr}{\mathsf{ErbB3}}
\DeclareMathOperator{\ErbBfour}{\mathsf{ErbB4}}

\DeclareMathOperator{\EGF}{\mathsf{EGF}}
\DeclareMathOperator{\HRG}{\mathsf{HRG}}
\DeclareMathOperator{\EGFR}{\mathsf{EGFR}}

\DeclareMathOperator{\MAPK}{\mathsf{MAPK}}

\DeclareMathOperator{\ERK}{\mathsf{ERK}}

{\tiny }

\newcommand{\ovl}{\mathbin{\lhd\mkern-9mu-}}

\title{Towards Scalable Modeling of Biology in Event-B}
\author{Usman Sanwal \inst{1}\and
        Thai Son Hoang\inst{2}\and 
        Luigia Petre \inst{1}\and
        Ion Petre \inst{3,4}
        }
\institute{Faculty of Science and Engineering,
\AA bo Akademi University, Finland \and
School of Electronics and Computer Science,
University of Southampton, UK \and
   Department of Mathematics and Statistics,
   University of Turku, Finland 
\and National Institute for Research and Development in Biological Sciences, Romania
   }

\begin{document}

\maketitle
\begin{abstract}
%Complexity and size are both the reason it is difficult to model and analyze a biological system as well as why it is essential we are able to, to understand their mechanisms and rules. In this paper we demonstrate a proof-of-concept for better addressing complexity and size when modeling biological systems. We employ the common mathematical functions in Event-B, a state-based formal method that has refinement as its central ingredient. Using functions leads to a more elegant and concise modeling, allowing us to formally capture a detailed biological system - the ErbB signaling pathway - with 242 events (state changes) in Event-B. To the best of our knowledge, this is the largest Event-B model ever built.

Biology offers many examples of large-scale, complex, concurrent systems: many processes take place in parallel, compete on resources and influence each other's behavior. The scalable modeling of biological systems continues to be a very active field of research. In this paper we introduce a new approach based on Event-B, a state-based formal method with refinement as its central ingredient, allowing us to check for model consistency step-by-step in an automated way. Our approach based on functions leads to an elegant and concise modeling method. We demonstrate this approach by constructing what is, to our knowledge, the largest ever built Event-B model, describing the ErbB signaling pathway, a key evolutionary pathway with a significant role in development and in many types of cancer. The Event-B model for the ErbB pathway describes $1320$ molecular reactions through $242$ events. 
\end{abstract}

\section{Introduction}\label{sec:intro}
Biological systems are typically very large and complex, so much that it is remarkably difficult to capture all the necessary details in one modeling step. The concept of refinement -- gradually adding details to a model while preserving its consistency -- is thus instrumental and it has been shown before, in our~\cite{Usman_2017} and related research~\cite{danos2009rule,quantrefFour,gratie14hiding,iancu2012quantitative}, to bring value to biological modeling.

Biological systems are often modeled by so-called reaction networks, i.e., as sets of biochemical reactions of type \[n_A*A+n_B*B\rightleftarrows n_C*C+n_D*D,\] 
where the reactants $A$, $B$, $C$, and $D$ model species, proteins, genes, etc. and $n_A$, $n_B$, $n_C$, $n_D \in \mathbb{N}$. When refining a reaction network, usually more reactants and their corresponding new reactions are added, and/or some (abstract) reactions are replaced with specialized sets of reactions, more accurately modeling the phenomenon of interest. In both cases, the reaction networks grow, sometimes exponentially; suitable tools for handling the models, their correctness properties, as well as their refinements are needed.

Reaction network modeling, including refinement, has already been addressed with different approaches, such as ODE-based modeling \cite{iancu2012quantitative}, rule-based modeling \cite{danos2009rule}, Petri nets \cite{gratie14hiding}, guarded command languages \cite{quantrefFour} and Event-B~\cite{Usman_2017}. Event-B~\cite{Abrial:2010}, in particular, is especially suitable for modeling complex systems, due to the concept of stepwise refinement that is a central part of this formal method. New details of a model are introduced by adding new variables (that model the state of the system) and events (that model the state changes in the system), potentially in several different refinement steps. This makes the modeling of large and complex systems more manageable. Correctness properties that were proved for a model are preserved when refining that model: the refinement approach is also called correctness-by-construction.
%Event-B has an associated tool support in the form of the Rodin platform (Rodin)~\cite{AbrialBHHMV10} - an Eclipse-based toolset for formally modeling and analyzing systems. To prove any property (e.g., invariance, termination, or refinement), Rodin generates all the necessary proof obligations (POs) that must be discharged to ensure the consistency of the model. Then, the POs can be proved either automatically or interactively with input from the modeler, for instance by adding useful assumptions or choosing a different proof strategy. 
The advantage that Event-B brings, when compared to other approaches (Petri nets, ODE modeling, etc.) is that it has refinement as the key concept of the development method and is supported by a toolset named Rodin~\cite{AbrialBHHMV10}. System details can be introduced in several steps and the tool manages all the links between all the intermediary models. Consistency of refinement ensures that all the properties of a model $M_i$ are still valid in its direct refinement successor $M_{i+1}$. At each refinement step, one can focus on the new elements that are introduced and on their consistency with the previous model. This approach allows to also separate the reasoning about the system under development into smaller steps.

In this paper, we model two biological systems using refinement in Event-B, i.e., we first model a simple, more abstract model of the system and then we add more details in a correct-by-construction manner, as explained above. The two systems we address are the heat shock response and the $\ErbB$ signaling pathway. Modeling the heat shock response in Event-B succeeded before~\cite{Usman_2017}: we started with the abstract model having $10$ variables and $17$ events and ended up with the concrete model having $22$ variables and $57$ events. Modeling the $\ErbB$ signaling pathway only succeeded earlier~\cite{IANCU2019} for the abstract model, with $110$ variables and $242$ events. The concrete model would have $1320$ events, which was not supported by Rodin.

The contribution of this paper consists in demonstrating how a particular modeling feature of Event-B -- the common mathematical function -- enables us to significantly reduce the concrete models' sizes. The relation between the abstract and the concrete forms of a reactant is captured with a function. This enables us to model the concrete reactions more elegantly and concisely, and as a result, the total number of events in the refined model is reduced significantly. In the case of the heat shock response, the complete model is described through $21$ events, instead of the $57$ events of the model in \cite{Usman_2017}. The difference in the case of the $\ErbB$ model is drastic, as we now need only 242 events for the full model of the $\ErbB$ signaling pathway in Rodin, instead of $1320$ events. Rodin is successfully handling this.

Thus, based on our experiments with modeling the two biological systems, we demonstrate a proof-of-concept about employing functions to address scalability with Event-B. To the best of our knowledge, the concrete $\ErbB$ signaling pathway model with $242$ events is the biggest Event-B model ever built. This is significant, since we now have proof of how to manage the modeling and analysis of large systems formally.

We proceed as follows. In Section~\ref{sec:prelim} we review the biological systems we address (the heat shock response and the $\ErbB$ signaling pathway); we also discuss Event-B particulars. In Section~\ref{sec:MNEventB}, we present our scheme for building an Event-B model corresponding to a given reaction network, also introducing the function-based modeling idea. In Sections~\ref{sec:HSRModel} and~\ref{sec:ErbBModel}, we illustrate the function-based modeling in Event-B of the heat shock response and of the $\ErbB$ signaling pathway, respectively. We discuss our results and potential impact in Section~\ref{sec:discussion}. All Event-B models constructed in the paper can be downloaded at \url{https://combio.org/wp-content/uploads/2021/05/Event-B_Model_ICTAC2021.zip}. 
\section{Preliminaries}\label{sec:prelim}
In this section we describe the biological systems we model -- the heat shock response and the $\ErbB$ signaling pathway -- and then we briefly review Event-B, the modeling method we use.

\subsection{The heat shock response (HSR)}\label{sec_hsr}

The heat-shock response is a cellular-level regulatory mechanism \cite{Powers20073758,PMID:7734836}. Proteins are folded in three dimensional shapes and the fold determines whether it can achieve its functionality (e.g., bind to a certain site on a DNA molecule or on another protein). Protein folding is a dynamical process, continuously influenced by many factors, such as chemical modifications of the amino-acids forming the protein (e.g., phosphorylation, acetylation, sumoylation) and properties of the environment (e.g., temperature, radiation, heavy metals). Misfolded proteins quickly form large protein bundles that are detrimental to the normal physiology of a cell and eventually lead to cell death. The heat shock response is one of the stress response mechanisms of a cell, aiming to limit the accumulation of misfolded proteins and assisting misfolded proteins to regain their natural fold. The heat shock response synthesizes a group of proteins -- called heat shock proteins ($\hsp$s) -- that act as molecular chaperones for the misfolded proteins and support their recovery from stress.  This is achieved either by repairing the damaged proteins or by degrading them, thus restoring protein homeostasis and promoting cell survival. Without such a mechanism, misfolded proteins will form plaque, which is the hallmark of many neurological diseases.

\begin{table}
	%{\small
	\begin{center}
		\caption{The molecular model for the eukaryotic heat shock response proposed in ~\cite{petre2011simple}.}\label{hsr_model_table}
		\begin{tabular}{rl@{\quad\quad}rl}
			\hline
			(1) &   $2 \hsf \rightleftarrows\hsfdimer$ & (7)    &$\hsp+\hsftrimer \rightarrow \hsphsf +2 \hsf$\\
			(2) &   $\hsf +\hsfdimer \rightleftarrows\hsftrimer$ & (8) &$\hsp+\hsfhse \rightarrow \hsphsf +2 \hsf +\hse$\\
			(3) &   $\hsftrimer +\hse \rightleftarrows\hsfhse$  &(9) &$\hsp \rightarrow \emptyset$\\
			(4) &   $\hsfhse \rightarrow \hsfhse + \hsp$ &  (10)     &$\prot \rightarrow \mfp$\\
			(5) &   $\hsp +\hsf \rightleftarrows\hsphsf$ & (11)  &$\hsp+\mfp \rightleftarrows\hspmfp$\\
			(6) &   $\hsp+\hsfdimer \rightarrow \hsphsf + \hsf$ & (12) &$\hspmfp \rightarrow \hsp+\prot$\\
			\hline
		\end{tabular}
	\end{center}
	%}
\end{table}

The basic model we discuss for the eukaryotic heat shock response is presented in \cite{petre2011simple} and summarized in Table \ref{hsr_model_table}. When the temperature increases, proteins $\prot$ begin to misfold, namely transform into $\mfp$ (Reaction (10)). The heat shock proteins have a high affinity to bind to the misfolded proteins, acting as chaperones and forming $\hspmfp$ complexes (Reaction (11)). Then, the complex $\hspmfp$ can transform back into the original protein $\prot$, freeing the heat shock factor protein $\hsp$ too (Reaction (12)). The $\hsp$ is synthesized as follows. A specific (called transcription factor) protein -- known as the heat shock factor ($\hsf$) -- binds in trimmer form to the $\hsp$'s gene promoter -- the heat shock element $\hse$ (Reactions (1)--(3) in Table \ref{hsr_model_table}). The formed $\hsfhse$ then produces the $\hsp$ proteins (Reaction (4)). These tend to combine with $\hsf$ and stay in inactive state as $\hsphsf$ complexes (right arrow in Reaction (5), as well as Reactions (6)--(8)). Once the temperature increases and more $\hsp$ are becoming chaperons for $\mfp$, less are available for forming $\hsphsf$ complexes and the balance changes: the left arrow in the Reaction (5) is activated. Finally, $\hsp$s can also degrade (Reaction (9)).

This is a simplified description of heat shock response, which is much more complex. As an example of the complexity, proteins can have multiple forms once they are synthesized, for instance they can be phosphorylated (slightly altered by an enzymatic reaction with an extra $PO_4$ phosphate group; since $PO_4$ has a negative electrical charge, this means that the protein folding is slightly different, leading to a changed activity). In this paper, we focus on the phosphorylation of only one aminoacid -- called S230 -- of the $\hsf$ protein. In our more detailed model, we take into account two versions of $\hsf$: one where S230 is present in the non-phosphorylated form (denoted $\rhsfzero$) and the other where S230 is present in the phosphorylated form (denoted $\rhsfone$). The full details for the refinement of the heat shock response can be found in \cite{petre2011simple,Usman_2017}.

\subsection{The $\ErbB$ signaling pathway}\label{sec_erbb}

The $\ErbB$ signaling pathway is a very well studied evolutionary pathway, because it is essential in the growth and expansion of organs and of the central nervous system~\cite{Oda:2005aa,Birtwistle:2007aa,Chen:2009aa}. Its main role is to induce, through the cellular membrane, a signal instigating the cell's growth and differentiation. This pathway is often overly active in various types of cancer and has been used for a long time as a therapeutic target. Once activated, the pathway keeps signaling to the cell to grow and differentiate, potentially leading to the uncontrolled growth that is the hallmark of cancer. 

We discuss briefly here the key functionality of the $\ErbB$ signaling pathway using a highly simplified language. For details we refer to \cite{hornberg2005control,Kholodenko:1999aa,Schoeberl:2002aa}. The epidermal growth factors ($\EGF$) are a family of proteins that signal to cells to grow and differentiate. They do that by binding to ligand proteins embedded in the cellular membrane -- the epidermal growth factor receptors ($\EGFR$). Once bound, the complex dimerizes and then gets phosphorylated. This then activates other ($\MAPK$ and $\ERK$) signaling pathways. All of these activations are done step by step through a cascade of reactions, whose effect is the activation of some proteins, that then participate in other reactions activating other proteins, etc. 

We follow in this paper the model of the $\ErbB$ signaling pathway presented in~\cite{hornberg2005control}, that is a revised version of the two earlier models presented in~\cite{Kholodenko:1999aa} and~\cite{Schoeberl:2002aa}. The model is first presented on a more generic level, along the lines briefly described above. This initial model consists however of $148$ reactions. The full model is then introduced essentially by differentiating between the four members of the $\EGFR$ family ($\ErbBone$ (also known as $\EGFR$), $\ErbBtwo$, $\ErbBthr$, $\ErbBfour$) and the two members of the $\EGF$ family ($\EGF$ and $\HRG$). Adding these details leads to many more species in the model. For example, the bonded complex $\EGF:\EGFR$ is replaced by 8 different variants of it, and the dimer $(\EGF:\EGFR)_2$ is replaced by 64 complexes. The full model of~\cite{hornberg2005control} has $1320$ reactions.

\subsection{Event-B}\label{subsec:EB}

\begin{figure}[htb]
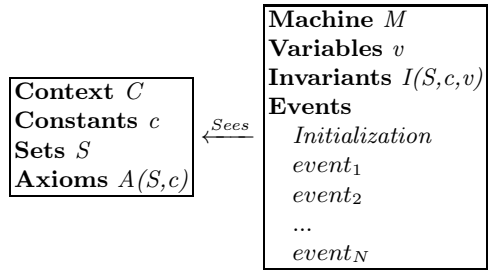

	\begin{center}
		\begin{tabular}{lll}
			
			\begin{tabular}{|l|}
				\hline
				\bf{Context} \it{C}\\
				\bf{Constants} \it{c}\\
				\bf{Sets} \it{S}\\
				\bf{Axioms} \it{A(S,c)}\\
				\hline
			\end{tabular}
			
			&
			
			$\xleftarrow{\text{\it{Sees}}}$
			
			&
			
			\begin{tabular}{|l|}
				\hline
				\bf{Machine} \it{M}\\
				\bf{Variables} \it{v}\\
				\bf{Invariants} \it{I(S,c,v)}\\
				\bf{Events}\\
				\ \ \ \it{Initialization}\\
				\ \ \ \it{$event_1$}\\
				\ \ \ \it{$event_2$}\\
				\ \ \ ...\\
				\ \ \ \it{$event_N$}\\
				\hline
			\end{tabular}

		\end{tabular}
	\end{center}
	\caption{General structure of machine \textit{M} and context \textit{C} in Event-B}\label{Gen_rep_table}
\end{figure}

Event-B~\cite{Abrial:2010} is a state-based formal method, building on earlier formalisms such as the B-Method~\cite{Abrial:1996} and the Action Systems~\cite{Back:1983}. The system state in Event-B is described by the values of \emph{variables} and the state changes are modeled using \emph{events}. The types of variables and other important properties that must hold during system execution are defined as \emph{invariants}. The initial system state is described with a specific event named \emph{Initialisation}. An event can contain \emph{parameters}, a \emph{guard} and an \emph{action}. The parameters model some variables local to the event; the guard is a predicate on the variables and parameters, describing the conditions under which the action can take place; the action describes the updates to the variables. If a guard evaluates to true, then we say that the event is \emph{enabled}. If two or more events are enabled at the same time, then one is non-deterministically chosen and executed.
%If two events do not update each other's variables, then they can be executed in any order and we can consider that they are executed in parallel.
The variables and events in an Event-B model of a system are contained in a \emph{machine}, also referred to as the ``dynamic part'' of the model. An Event-B machine can \emph{see} one or more \emph{contexts}, also known as the ``static part'' of the model. A context contains definitions of constants, carrier sets, as well as axioms about them. A general structure of an Event-B model, made out of machine $M$ and context $C$ is presented in Fig.~\ref{Gen_rep_table}.

A key concept in formal modeling with Event-B is that of \emph{refinement}~\cite{Abrial:2010}: this allows the modeler to start from a simple model of the system and then gradually introduce more details, in the form of new events, variables or context data.

Event-B modeling has two types of refinements: \textit{superposition refinement} and \textit{data refinement}. Superposition refinement~\cite{Back1996,Katz:1993} is the term used when we refine a model by adding new variables and events to the existing model. The validity of model is preserved by making sure that newly added events must neither contradict nor take over the previous events in any of the preceding models. In \emph{data refinement}~\cite{HOARE198771},  some variables in the more abstract machine are replaced by other variables in the refined machine; in this case, we need to add a \emph{gluing invariant} in the refined machine, which formally defines the relation between the previous, abstract variables and the newly introduced, concrete ones. Refinement in Event-B  has been used to model numerous protocols and systems, see~\cite{Abrial:2010,Butler2007,Lanoix2008,Hoang2009,kamali2010,SalehiFathabadi2011,petre2012,horsmanheimo2014,kamali2015,Usman_2017}.

Event-B benefits from the tool support of the Eclipse-based Rodin platform~\cite{AbrialBHHMV10}. Rodin allows to edit the model, to prove properties of the model, to animate the model and even allows model checking. Proving in Event-B employs several proof engines to automatically prove the different properties of the model. This works by Rodin automatically generating first proof obligations in the form of sequents; these need to be discharged in order for the different properties (e.g., invariance, termination, or refinement) to hold. The automatic provers usually discharge many of the proof obligations. The remaining ones can be tackled using the interactive prover, with input from the modeler, for instance by adding useful assumptions or choosing a different proof strategy. The fact that some properties are not discharged automatically shows that there might be some problem with some modeling aspect of the system. The modeler then has a chance to edit the model to address the issue. Such interleaving between modeling and proving is an important aspect of working with the Rodin platform and is quite similar to the compilation of programs~\cite{AbrialBHHMV10}.

\section{Modelling reaction networks in Event-B}\label{sec:MNEventB}

%\begin{itemize}
% \item Our older approach, discussed for the heat shock response. Very brief but still clear enough so the paper is self-contained.
%\item One event per reaction
%\item Invariants for mass-conservation
%\item Refinement with glue invariants for the refinement relation.	
%\item Our new approach: the use of functions. */
%\end{itemize}

We model reaction networks as sets of biochemical reactions, where each reaction specifies its reactants, products, and possibly inhibitors and catalyzers, see \cite{Klipp-book}. These reactions can be either reversible or irreversible and each reaction could also have an associated flux, describing the rate at which its products are produced and its reactants consumed.  For simplicity, we consider each reversible reaction in our methodology as two reactions and we model only the reactants and the products in this paper.
%also assume non-enzymatic reactions only.
With these assumptions, a reaction $r$ can be written as a rewriting rule of the form:

\vspace*{-0.2cm}
\begin{equation}\label{eq_R}\tag{R}
r: \ \ \ m_1X_1 + m_2X_2 +...+ m_nX_n \rightarrow m'_1X_1 + m'_2X_2 +...+ m'_nX_n,
\end{equation}

\noindent where  ${\cal S} = \{X_1,\ldots,X_n\}$ is the set of \emph{reactants} and $m_1,..., m_n, m'_1,..., m'_n \in \mathbb{N}$ are non-negative integers.

Reaction networks are modeled rather straightforwardly in Event-B: every reactant is modeled by a variable and every reaction is modeled by an event. Invariants ensure the correctness of each reactant modeling as well as other biological properties of interest, for instance the mass conservation rule that requires that the number of certain reactants is constant in the system. 

Thus, $X_1, X_2$ $,...,X_n$ are the variables of the model, their type (the set of non-negative integers) being specified by corresponding invariants. This is another simplification through which we consider each species as being discrete instead of continuous. Initial values for all of these variables are set in the initialisation event. For each reaction $r$ of the reaction network, we specify in its guard that it must have enough of each reactant in order for the reaction to be enabled, while the action of the event specifies the changes to happen to each variable. The general form of an Event-B model corresponding to a reaction network as described by \eqref{eq_R} is presented in Table \ref{tab:general_model}. For more details of this general scheme, we refer to \cite{Usman_2017}.

\vspace*{-0.2cm}
\begin{table}[htb]
	%{\small
	\begin{center}
		\caption{The general form of an Event-B model for a reaction network.}\label{tab:general_model}
		
		\begin{tabular}[htb]{p{6cm} p{6cm}}
			
			\begin{tabular}{l}

				\begin{tabular}{|L{5cm}|}
					\hline
					{\bf VARIABLES} $X_1, X_2,...,X_n$ \\
					{\bf INVARIANTS} \\
					\ \ {\bf @inv1} $X_1\in \mathbb{N}$\\
					\ \ {\bf @inv2} $X_2\in \mathbb{N}$\\
					\ \ $\ldots$\\
					\ \ {\bf @invn} $X_n\in \mathbb{N}$\\
					\hline
				\end{tabular}
				
				\\
				\\

				\begin{tabular}{|L{5cm}|}
					\hline
					{\bf INITIALISATION} \\
					%{\bf THEN} \\
					\ \ {\bf @act1} $X_1 = init_1$\\
					\ \ {\bf @act2} $X_2 = init_2$\\
					\ \ $\ldots$\\
					\ \ {\bf @actn} $X_n = init_n$\\
					\hline
				\end{tabular}
				
			\end{tabular}
			
			&
			
			\begin{tabular}{|L{5cm}|}
				\hline
				{\bf Event r} \\
				{\bf WHERE} \\
				\ \ {\bf @grd1} $X_1 \geq m_1$\\
				\ \ {\bf @grd2} $X_2 \geq m_2$\\
				\ \ $\ldots$\\
				\ \ {\bf @grdn} $X_n \geq m_n$\\
				{\bf THEN}  \\
				\ \ {\bf @act1} $X_1 := X_1 + (m'_1 - m_1)$\\
				\ \ {\bf @act2} $X_2 := X_2 + (m'_2 - m_2)$\\
				\ \ $\ldots$\\
				\ \ {\bf @actn} $X_n := X_n + (m'_n - m_n)$\\
				{\bf END}\\
				\hline
			\end{tabular}
			
		\end{tabular}
	\end{center}
	%}
\end{table}

\vspace*{-1.2cm}
In this work we use data refinement to add detail to a biological model. In the refined machine, we introduce gluing invariants modeling the relations between the variables-to-be-refined from the abstract machine and their concrete versions in the refined machine. In the following, we present the differences with respect to scalability in two different types of reactions, a binding and a dimerization. In Sections~\ref{sec:HSRModel} and~\ref{sec:ErbBModel}, we point out how this type of modeling was used for the heat shock response and the $\ErbB$ signaling pathway, respectively.

\subsection{Binding}
%\label{bind}

Say we have a reaction (bind) of type
$A + B \rightarrow AB$. According to Table \ref{tab:general_model}, the corresponding event is shown in event {\bf AbstractBind} (bind$\_$EB). Further assume the $A$ reactant is to be refined into two special cases, $A_0$ and $A_1$. The less scalable data refinement approach, used in~\cite{Usman_2017}, leads us to refine the (bind$\_$EB) event into the two events shown in Table~\ref{tab:bindtrad}, where $A_0B$ and $A_1B$ are the refined bindings and the gluing invariants are $A=A_0+A_1$ and $AB=A_0B + A_1B$.

\[\tag{bind\_EB}	{\small\begin{tabular}{|L{5cm}|}
				\hline
				{\bf AbstractBind event}\\
				{\bf WHERE} \\
				\ \ {\bf @grd1} $A \geq 1$\\
				\ \ {\bf @grd2} $B \geq 1$\\
				{\bf THEN}  \\
				\ \ {\bf @act1} $A := A - 1$\\
				\ \ {\bf @act2} $B := B - 1$\\
				\ \ {\bf @act3} $AB := AB + 1$\\
				{\bf END}\\
				\hline
			\end{tabular}.}\]

\vspace*{-0.2cm}
\begin{table}[htb]
	{\small
	\begin{center}
		\caption{The traditional binding data refinement approach}\label{tab:bindtrad}
		
		\begin{tabular}{p{6cm} p{6cm}}
			
			\begin{tabular}{l}

				\begin{tabular}{|L{5cm}|}
					\hline
					{\bf ConcreteBind0 event} \\
				{\bf WHERE} \\
				\ \ {\bf @grd1} $A_0 \geq 1$\\
				\ \ {\bf @grd2} $B \geq 1$\\
				{\bf THEN}  \\
				\ \ {\bf @act1} $A_0 := A_0-1$\\
				\ \ {\bf @act2} $B := B-1$\\
				\ \ {\bf @act3} $A_0B := A_0B +1$\\
				{\bf END}\\
					\hline
				\end{tabular}
				
			\end{tabular}
			
			&
			
			\begin{tabular}{|L{5cm}|}
				\hline
				{\bf ConcreteBind1 event} \\
				{\bf WHERE} \\
				\ \ {\bf @grd1} $A_1 \geq 1$\\
				\ \ {\bf @grd2} $B \geq 1$\\
				{\bf THEN}  \\
				\ \ {\bf @act1} $A_1 := A_1-1$\\
				\ \ {\bf @act2} $B := B-1$\\
				\ \ {\bf @act3} $A_1B := A_1B +1$\\
				{\bf END}\\
				\hline
			\end{tabular}
			
		\end{tabular}
	\end{center}
	}
\end{table}

\vspace*{-1.9cm}
\begin{table}[htb]
		\caption{The scalable binding data refinement approach}\label{tab:bindscal}
	{\small
		\begin{center}
		\begin{tabular}{p{6cm} p{6cm}}

\vspace*{-0.9cm}
				\begin{tabular}{|l|}
					\hline
						\bf{Context...} \\
				\bf{Constants}
				    \hspace{2mm}$A_0, A_1, A_0B, A_1B$\\
				\bf{Sets}
				    \hspace{2mm}$A\_SET, AB\_SET$\\
				\bf{Axioms} \\
				    %\hspace{2mm}$A_0 \neq A_1$\\
				   %\hspace{2mm}$A_0B \neq A_1B$\\
				   \hspace{2mm}partition($A\_SET, \{A_0\}, \{A_1\})$\\
				   \hspace{2mm}partition($AB\_SET, \{A_0B\}, \{A_1B\})$\\[1ex]
					\hline
				\end{tabular}

			%& 
			
			\begin{tabular}{|l|}
				\hline
				{\bf ScalableConcreteBind event} \\
				{\bf ANY} $e,i$\\
				{\bf WHERE} \\
				\ \ {\bf @grd1} $A\_FUNC(e) \geq 1$\\
				\ \ {\bf @grd2} $B \geq 1$\\
				\ \ {\bf @grd3} $(e=A_0 \wedge i=A_0B) \vee (e=A_1 \wedge i=A_1B)$\\
				{\bf THEN}  \\
				\ \ {\bf @act1} $A\_FUNC(e) :=A\_FUNC(e)-1$\\
				\ \ {\bf @act2} $B := B-1$\\
				\ \ {\bf @act3} $AB\_FUNC(i) :=AB\_FUNC(i) +1$\\
				{\bf END}\\
				\hline
			\end{tabular}
			
		\end{tabular}
		\end{center}
	}
\end{table}

\vspace*{-1.2cm}
If, instead of defining four concrete variables $A_0, A_1, A_0B, A_1B$ in the refined machine as non-negative integers, we define two functions whose domains are special constant sets defined in the context, then we can replace the two events {\bf ConcreteBind0} and {\bf ConcreteBind1} with only one event, named {\bf ScalableConcreteBind}, illustrated in Table~\ref{tab:bindscal}. The two concrete variables are now $A\_FUNC$ and $AB\_FUNC$, defined (in invariants in the refined machine) as functions $A\_FUNC: A\_SET \rightarrow \mathbb{N}$ and $AB\_FUNC: AB\_SET \rightarrow \mathbb{N}$, respectively. The gluing invariants are $A = A\_FUNC(A_0) + A\_FUNC(A_1)$ and
$AB = AB\_FUNC(A_0B) + AB\_FUNC(A_1B)$.

\subsection{Dimerization}
%\label{bind}

Say we have a reaction (dimer) of type
$A + A \rightarrow AA$. According to Table \ref{tab:general_model}, its corresponding event is

\vspace*{-0.2cm}
\[\tag{dimer\_EB}\begin{tabular}{|L{5cm}|}
				\hline
				{\bf AbstractDimer event}\\
				{\bf WHERE} \\
				\ \ {\bf @grd1} $A \geq 2$\\
				{\bf THEN}  \\
				\ \ {\bf @act1} $A := A - 2$\\
				\ \ {\bf @act3} $AA := AA + 1$\\
				{\bf END}\\
				\hline
			\end{tabular}.\]

\noindent Further assume the $A$ reactant is to be refined into two special cases, $A_0$ and $A_1$. In the data refinement approach in~\cite{Usman_2017}, we refine the (dimer$\_$EB) event into three events (see Table~\ref{tab:dimertrad}), with the refined dimers $AA_0$, $AA_1$ and $AA_{01}$ and the gluing invariants $A = A_0 + A_1$ and
$AA = AA_0 + AA_1+AA_{01}$.

\vspace*{-0.5cm}
\begin{table}[htb]
	{\small
	\begin{center}
		\caption{The traditional dimer data refinement approach}\label{tab:dimertrad}
		
		\vspace*{-0.5cm}
		\begin{tabular}{ccc}
			
			%\begin{tabular}{l}

				\begin{tabular}{|L{3.8cm}|}
					\hline
					{\bf ConcreteDimer0 event} \\
				{\bf WHERE} \\
				\ \ {\bf @grd1} $A_0 \geq 2$\\
				{\bf THEN}  \\
				\ \ {\bf @act1} $A_0 := A_0-2$\\
				\ \ {\bf @act3} $AA_0 := AA_0 +1$\\
				{\bf END}\\
					\hline
				\end{tabular}
				
			%\end{tabular}
			
			&
			
			\begin{tabular}{|L{3.8cm}|}
				\hline
				{\bf ConcreteDimer1 event} \\
				{\bf WHERE} \\
				\ \ {\bf @grd1} $A_1 \geq 2$\\
				{\bf THEN}  \\
				\ \ {\bf @act1} $A_1 := A_1-2$\\
				\ \ {\bf @act3} $AA_1 := AA_1 +1$\\
				{\bf END}\\
				\hline
			\end{tabular} &
			
			\begin{tabular}{|L{4cm}|}
				\hline
				{\bf ConcreteDimer01 event} \\
				{\bf WHERE} \\
				\ \ {\bf @grd1} $A_0 \geq 1$\\
				\ \ {\bf @grd2} $A_1 \geq 1$\\
				{\bf THEN}  \\
				\ \ {\bf @act1} $A_0 := A_0-1$\\
				\ \ {\bf @act2} $A_1 := A_1-1$\\
				\ \ {\bf @act3} $AA_{01} :=AA_{01} +1$\\
				{\bf END}\\
				\hline
			\end{tabular}\\
			
		\end{tabular}
	\end{center}
	}
\end{table}

\vspace*{-1.2cm}
If, instead of defining five concrete variables $A_0, A_1, AA_0$, $AA_1$, $AA_{01}$ in the refined machine as non-negative integers, we define two functions whose domains are special constant sets defined in the context, then we can replace the three events {\bf ConcreteDimer0}, {\bf ConcreteDimer1} and {\bf ConcreteDimer01} with only two events, named {\bf ScalableConcreteBind} and {\bf ScalableConcreteBind01}, illustrated in Table~\ref{tab:dimerscal}. The concrete variables are now $A\_FUNC$ and $AB\_FUNC$, defined (in invariants in the refined machine) as functions $A\_FUNC: A\_SET \rightarrow \mathbb{N}$ and $AA\_FUNC: AA\_SET \rightarrow \mathbb{N}$, respectively. The gluing invariants are $A = A\_FUNC(A_0) + A\_FUNC(A_1)$ and
$AA = AA\_FUNC(AA_0) + AA\_FUNC(AA_1)+AA\_FUNC(AA_{01})$. The symbol $\ovl$ in Table~\ref{tab:dimerscal} is used so that the function $A\_FUNC$ is modified only for the elements $A_0$ and $A_1$ of its domain.

\begin{table}[htb]
	{\small
	\begin{center}
		\caption{The scalable dimer data refinement approach}\label{tab:dimerscal}
		
				\begin{tabular}{|L{6.8cm}|}
					\hline
						\bf{Context...} \\
				\bf{Constants} \\
				    \hspace{2mm}$A_0, A_1, AA_0, AA_1, AA_{01}$\\
				\bf{Sets} \\
				    \hspace{2mm}$A\_SET, AA\_SET$\\
				\bf{Axioms} \\
				 \hspace{2mm}partition($A\_SET,\{A_0\}, \{A_1\})$\\
                 \hspace{2mm}partition($AA\_SET, \{AA_0\}, \{AA_1\}, \{AA_{01}\})$\\
					\hline
				\end{tabular} \\[2ex]
				
			\begin{tabular}{|L{7.5cm}|}
				\hline
				{\bf ScalableConcreteDimer event} \\
				{\bf ANY} $e,i$\\
				{\bf WHERE} \\
				\ \ {\bf @grd1} $A\_FUNC(e) \geq 2$\\
				\ \ {\bf @grd2} $(e=A_0 \wedge i=AA_0)$ $\vee$ $(e=A_1 \wedge i=AA_1)$\\
				{\bf THEN}  \\
				\ \ {\bf @act1} $A\_FUNC(e) := A\_FUNC(e)-2$\\
				\ \ {\bf @act2} $AA\_FUNC(i) := AA\_FUNC(i) +1$\\
				{\bf END}\\
				\hline
			\end{tabular}
			\\[2ex]

			\begin{tabular}{|L{0.95\textwidth}|}
				\hline
				{\bf ScalableConcreteDimer01 event} \\
				{\bf ANY} $e,i$\\
				{\bf WHERE} \\
				\ \ {\bf @grd1} $A\_FUNC(A_0) \geq 1$\\
				\ \ {\bf @grd2} $A\_FUNC(A_1) \geq 1$\\
				{\bf THEN}  \\
				\ \ {\bf @act1} $A\_FUNC :=$
				$A\_FUNC$ $\ovl$ \\
				\ \ \hspace{3.5cm}$\{A_0 \mapsto A\_FUNC(A_0) - 1, A_1 \mapsto A\_FUNC(A_1) - 1\} $\\
				\ \ {\bf @act2} $AA\_FUNC(AA_{01}) :=$
				$AA\_FUNC(AA_{01}) +1$\\
				{\bf END}\\
				\hline
			\end{tabular}
			\\[2ex]
			%\end{tabular}
			
		%\end{tabular}
	\end{center}
	}
\end{table}

\vspace*{-0.2cm}
\subsection{In a nutshell}
\label{nut}
What we propose in this paper is to use as concrete variable functions instead of non-negative integers. In the context part of the model, we define a constant set for each variable-to-be-refined of the reaction network. This set contains all the refined forms of the variable-to-be-refined and is the domain of the function (concrete variable) in the refined machine. Since we can generalise the formulation of guards when we use functions, we do not need to distinguish between so many different cases, and hence, the number of events does not grow in a combinatorial explosion anymore. The number of variables remains constant.

\section{An Event-B model for the heat shock response using functions}\label{sec:HSRModel}

\begin{table}[htb]
	%{\small
	\begin{center}
		\caption{Basic model: the HSF sequestration  event.}\label{tab:HSF_Seq}
		\begin{tabular}{|l|c|l|} %|cllc|cllc|
			
			\cline{1-1}
			\textbf{HSF Sequestration Basic Event} \\
			\textbf{WHERE}\\
			\ \ \textbf{@grd1} $\hsp \geq 1 \wedge \hsf \geq 1$\\
			\textbf{THEN}\\
			\ \ \textbf{@act1} $\hsp := \hsp - 1$\\
			\ \ \textbf{@act2} $\hsf := \hsf - 1$\\
			\ \ \textbf{@act3} $\hsphsf := \hsphsf + 1$ \\
			\textbf{END}\\
			\cline{1-1}
		\end{tabular}
	\end{center}
	%}
\end{table}

Here we use the approach described in Section~\ref{sec:MNEventB} for modeling two reactions - (5) and (1) - of the heat shock response in Table~\ref{hsr_model_table} and their phosphorylation-related refinement. To model the refinement of the heat shock factor ($\hsf$) into its two variants (0- and 1-phosphorylated), we introduce a set named $\HSF$ and two distinct constants named $\HSFZ$ and $\HSFO$; $\HSF$ is defined as $\HSF=\{\HSFZ, \HSFO\}$. Similarly, $\hsf$ can be 0- or 1-phosphorylated also in the binding $\hsphsf$ with the heat shock protein $\hsp$. To capture this, we introduce a set named $\HSFHSP$ and two distinct constants named $\HSPFSFZ$ and $\HSPFSFO$; $\HSFHSP$ is defined as $\HSFHSP=\{\HSPFSFZ, \HSPFSFO \}$. Likewise, $\hsf$ can be 0- or 1-phosphorylated also in the dimer $\hsfdimer$. To capture this, we introduce the set $\RHSFdimer$ and three distinct constants $\HSFTZ$, $\HSFTO$, and $\HSFTT$, so that $\RHSFdimer = \{\HSFTZ, \HSFTO,\HSFTT\}$. This is implemented in the context part of the Event-B model. 

In~\cite{Usman_2017}, the abstract event in Table~\ref{tab:HSF_Seq} is replaced by two events (shown in Table~\ref{T_two_events}), where the concrete variables that replace $\hsf$ are $\rhsfzero$ and $\rhsfone$, with the gluing invariant $\hsf=\rhsfzero+\rhsfone$. Similarly, the abstract variable binding $\hsphsf$ is to be replaced by the two concrete variables $\hsprhsfnac$ and $\hsprhsfone$, with the gluing invariant $\hsphsf=\hsprhsfnac+\hsprhsfone$.

%WRITE HERE THE TWO EVENTS OF THE REFINED MODEL OF \cite{Usman_2017}

\vspace*{-0.3cm}
\begin{table}[htb]
	\begin{center}
		\caption{Previous approach: the refinement of the HSF sequestration  event. The variable $\hsf$ is replaced in all possible ways with $\rhsfzero$ and $\rhsfone$, leading to 2 events.}\label{T_two_events}
		
		\begin{tabular}{|l|c|l|} %|cllc|cllc|
			\cline{1-1}\cline{3-3}
			\textbf{HSF Sequestration Refinement-1}	&\hspace*{0.5cm}&\textbf{HSF Sequestration Refinement-2} \\
			\textbf{WHERE}				&&\textbf{WHERE}\\
			\ \ \textbf{@grd1} $\hsp \geq 1$ 	&&\ \ \textbf{@grd1} $\hsp \geq 1$\\
			\ \ \textbf{@grd2} $\rhsfzero \geq 1$ 	&&\ \
			\textbf{@grd2} $\rhsfone \geq 1$\\
			\textbf{THEN}					&&\textbf{THEN}\\
			\ \ \textbf{@act1} $\hsp := \hsp - 1$	&&\ \ \textbf{@act1} $\hsp := \hsp - 1$\\
			\ \ \textbf{@act2} $\rhsfzero := \rhsfzero - 1$	&&\ \ \textbf{@act2} $\rhsfone := \rhsfone - 1$\\
			\ \ \textbf{@act3} $\hsprhsfnac := \hsprhsfnac + 1$	&&\ \ \textbf{@act3} $\hsprhsfone := \hsprhsfone + 1$\\
			\textbf{END}					&&\textbf{END}\\
			\cline{1-1}\cline{3-3}
		\end{tabular}
	\end{center}
\end{table}

\vspace*{-1cm}
Now, instead of the concrete variables $\rhsfzero, \rhsfone, \hsprhsfnac, \hsprhsfone$ we define two functions, $\rhsf: \HSF \rightarrow \mathbb{N}$ and  $\rhsphsf: \HSFHSP \rightarrow \mathbb{N}$, so that we have the gluing invariants $\hsf = \rhsf(\HSFZ) + \rhsf(\HSFO)$ and $\hsphsf = \rhsphsf(\HSPFSFZ) + \rhsphsf(\HSPFSFO)$.
The event of the abstract model (Table~\ref{tab:HSF_Seq}) will be refined to a single event covering all the cases. So, instead of the two events shown in Table \ref{T_two_events}, we now have the single event in Table \ref{tab:HSF_Seq_fun}.

\begin{table}[htb]
	%{\small
	\begin{center}
		\caption{Current approach: the refinement of the HSF sequestration event. Functions are used for a compact formulation of the refinement.}\label{tab:HSF_Seq_fun}
		\begin{tabular}{|l|c|l|} %|cllc|cllc|
			
			\cline{1-1}
			\textbf{HSF Sequestration Refinement Functions} \\
			\textbf{ANY}\\
			\ \  $e, i$\\
			\textbf{WHERE}\\
			\ \ \textbf{@grd1} $\hsp \geq 1 \wedge \rhsfe \geq 1$\\
			\ \ \textbf{@grd2} $(e = \HSFZ  \land  \ i = \HSPFSFZ)$  $\lor$
			%\hspace{1.3cm}	
			$(e = \HSFO  \land \  i = \HSPFSFO) $ \\
			\textbf{THEN}\\
			\ \ \textbf{@act1} $\hsp := \hsp - 1$\\
			\ \ \textbf{@act2} $\rhsfe := \rhsfe - 1$\\
			\ \ \textbf{@act3} $\rhsphsf(i) := \rhsphsf(i) + 1$ \\
			\textbf{END}\\
			\cline{1-1}
		\end{tabular}
	\end{center}
	%}
\end{table}

%\vspace{-0.1cm}
\begin{table}[htb]
	\begin{center}
		\caption{Basic model: the HSF dimerization event.}\label{tab:HSF_dimer}
		\begin{tabular}{|l|c|l|} %|cllc|cllc|
			
			\cline{1-1}
			\textbf{Dimerization Basic Event} \\
			\textbf{WHERE}\\
			\ \ \textbf{@grd1} $\hsf \geq 2 $\\
			\textbf{THEN}\\
			\ \ \textbf{@act1} $\hsf := \hsf - 2$\\
			\ \ \textbf{@act2} $\hsfdimer := \hsfdimer + 1$\\
			
			\textbf{END}\\
			\cline{1-1}
		\end{tabular}
	\end{center}
	%}
\end{table}

%\vspace*{-0.1cm}
\begin{table}[htb]
	%{\small
	\begin{center}
		
		\vspace*{-0.1cm}
		\caption{Previous approach: the refinement of the HSF dimerization event. The variable $\hsf$ is replaced in all possible ways with $\rhsfzero$ and $\rhsfone$, leading to 3 events.}\label{tab:HSF_model}
		
		\vspace*{-0.1cm}
		\begin{tabular}{|l|c|l|} %|cllc|cllc|
			\cline{1-1}\cline{3-3}
			\textbf{Dimerization Refinement-1}&\hspace*{0.5cm}&\textbf{Dimerization Refinement-2}\\
			\textbf{WHERE}&&\textbf{WHERE}\\
			\ \ \textbf{@grd1} $\rhsfzero \geq 2$ && \ \  \textbf{@grd1} $\rhsfone \geq 2$\\
			\textbf{THEN}&& \textbf{THEN} \\
			\ \ \textbf{@act1} $\rhsfzero := \rhsfzero - 2$&& \ \ \textbf{@act1} $\rhsfone := \rhsfone - 2$\\
			\ \ \textbf{@act2} $\rhsfdimnac := \rhsfdimnac + 1$&& \ \ \textbf{@act2} $\rhsfdimactwo := \rhsfdimactwo + 1$\\
			\textbf{END}&&\textbf{END}\\
			\cline{1-1}\cline{3-3}\multicolumn{3}{c}{}
			\\
			\cline{1-1}
			\textbf{Dimerization Refinement-3} \\
			\textbf{WHERE}\\
			\ \ \textbf{@grd1} $\rhsfzero \geq 1 \wedge \rhsfone \geq 1$\\
			\textbf{THEN}\\
			\ \ \textbf{@act1} $\rhsfzero := \rhsfzero - 1$\\
			\ \ \textbf{@act2} $\rhsfone := \rhsfone - 1$\\
			\ \ \textbf{@act3} $\rhsfdimacone := \rhsfdimacone + 1$ \\
			\textbf{END}\\
			\cline{1-1}
		\end{tabular}
	\end{center}
\end{table}

%WRITE HERE THE 2 EVENTS, SYMMETRIC AND ASYMMETRIC (FROM THE MACHINE WITH A SINGLE STEP REFINEMENT)

%\vspace*{-1cm}
\begin{table}[htb]
	
	\begin{center}
		\caption{Current approach: the refinement of the HSF dimerization.}\label{Two_events_new}
		
		\begin{tabular}[t]{p{5.9cm}p{6.1cm}}
		
		    \begin{tabular}{|l|}
		    \hline \textbf{Dimerization Refinement Symetric}\\
		    \textbf{Any}\\
		    \ \ $e, f$ \\
		    \textbf{WHERE}\\
		    \ \ \textbf{@grd1} $\rhsfe \geq 2$\\
		    \ \ \textbf{@grd2} $(e = \HSFZ  \land  f = \HSFTZ) \lor$\\
		    \hspace{1.3cm}	$(e = \HSFO  \land  f = \HSFTT) $ \\
		    \textbf{THEN}\\
		    \ \ \textbf{@act1} $\rhsfe := \rhsfe - 2$\\
		    \ \ \textbf{@act2} $\rhsfdimerf := \rhsfdimerf + 1$\\
		    \textbf{END}\\
		    \hline
		    \end{tabular}
		    
		    &
		    
		    \begin{tabular}{|l|}
		    \hline
		    \textbf{Dimerization Refinement Asymetric} \\
            \textbf{WHERE}\\
			 \ \ \textbf{@grd1} $\rhsfZ \geq 1$\\
			 \ \ \textbf{@grd2} $\rhsfO \geq 1$\\
			\textbf{THEN}\\
			\ \ \textbf{@act1} $\rhsf := \rhsf\ovl$  \\
			\ \ \hspace{0.5cm}	$\{\HSFZ \mapsto \rhsf(\HSFZ) - 1,$\\
			\ \ \hspace{0.5cm}	$\HSFO \mapsto \rhsf(\HSFO) - 1 \}$\\
			\ \ \textbf{@act2} $\rhsfdimertwo :=$\\
			\ \ \hspace{0.5cm}$\rhsfdimertwo + 1$\\
			\textbf{END}\\
			\hline
		    \end{tabular}
	\end{tabular}
	\end{center}
\end{table}

%\vspace*{-1cm}
For $\hsf$'s dimerization (reaction (1) in Table~\ref{hsr_model_table}), we need to refine the abstract event in Table~\ref{tab:HSF_dimer}. In \cite{Usman_2017}, this event is refined by three events (shown in Table~\ref{tab:HSF_model}), where the concrete variables that replace $\hsf$ are $\rhsfzero$ and $\rhsfone$, with the gluing invariant $\hsf=\rhsfzero+\rhsfone$. Similarly, the abstract dimer variable $\hsfdimer$ is to be replaced by three concrete variables $\rhsfdimnac$, $\rhsfdimacone$ and $\rhsfdimactwo$, with the gluing invariant $\hsfdimer=\rhsfdimnac+\rhsfdimacone+ \rhsfdimactwo$.

Now, instead of the concrete variables $\rhsfzero, \rhsfone, \rhsfdimnac, \rhsfdimacone, \rhsfdimactwo$ we define two functions, $\rhsf: \HSF \rightarrow \mathbb{N}$ and  $\rhsfdimer: \RHSFdimer \rightarrow \mathbb{N}$, so that we have the gluing invariants $\hsf = \rhsf(\HSFZ) + \rhsf(\HSFO)$ and $\hsfdimer = \rhsfdimer(\HSFTZ) + \rhsfdimer(\HSFTO) + \rhsfdimer(\HSFTT)$.
The event of the abstract model (Table~\ref{tab:HSF_dimer}) will be refined by two events. So, instead of the three events shown in Table \ref{tab:HSF_model}, we now have two events in Table \ref{Two_events_new}.

Thus, our new approach leads to a new, more compact, refinement-based approach to biological modeling. In the case of the heat shock response, the complete model is described through  10 variables and  21 events, instead of the 22 variables and 57 events of the model in \cite{Usman_2017}. The full model can be downloaded at \url{https://combio.org/wp-content/uploads/2021/05/Event-B_Model_ICTAC2021.zip}. 

Also noteworthy is that, in \cite{Usman_2017}, if multiple variables need to be refined in one event, we refine one variable per refinement step; as a result, we refine the basic $\HSR$ model in 5 different refinement steps. Here, we refine all the variables of the event in one refinement step, since there are not so many new variables to handle and is conceptually clearer.

%\begin{itemize}
%\item Compare here the outcome of this approach with the outcome of the old approach. Compare number of variables and events. 	
%\end{itemize}

\section{An Event-B model for the $\ErbB$ signalling pathway using functions}\label{sec:ErbBModel}

%We built the basic model of $\ErbB$ in \cite{IANCU2019} and also checked its consistency. Here we insure the consistency of the refined model of $\ErbB$. We introduce the refinement of the $\ErbB$ signaling pathway model using functions. 

We extended the basic Event-B model of the $\ErbB$ signaling pathway presented in \cite{IANCU2019} to include details about epidermal growth factor receptor ($\EGFR$) and epidermal growth factor ($\EGF$). The epidermal growth factor receptor is refined into the four receptor members of the $\ErbB$ family: $\ErbBone$, $\ErbBtwo$, $\ErbBthr$, $\ErbBfour$. Also, the epidermal growth factor is refined into two types: $\EGF$ and $\HRG$. We refined all the reactions of the basic model of $\ErbB$ signaling pathway present in \cite{IANCU2019} where $\EGFR$ and $\EGF$ are present as a single species or present in the form of a dimer. This data refinement is presented as follows:
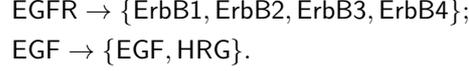
\begin{align*}
&\EGFR \rightarrow \{\ErbBone, \ErbBtwo, \ErbBthr, \ErbBfour\}; \\
&\EGF \rightarrow \{\EGF, \HRG\}.
\end{align*}
To refine the dimers, the two {sets} $\EGFEGFRrdim$ and $\EGFEGFRxrdim$ are each partitioned into eight, accounting for the eight possible forms of these dimers:

\begin{description}
	\item
	{\bf @axm3}:\\
	$partition(\EGFEGFRrdim$,$\{\EGFErbBonedim\}$, $\{\EGFErbBtwodim\}$,\\
	\hspace*{1.4cm} $\{\EGFErbBthrdim\}$,
	$\{\EGFErbBfourdim\}$,$\{\HRGErbBonedim\}$,\\
	\hspace*{1.4cm} $\{\HRGErbBtwodim\}$, $\{\HRGErbBthrdim\}$, $\{\HRGErbBfourdim\})$

	\item
	{\bf @axm4}:\\
	$partition(\EGFEGFRxrdim$,$\{\EGFErbBonexdim\}$, $\{\EGFErbBtwoxdim\}$, \\
	\hspace*{1.4cm} $\{\EGFErbBthrxdim\}$, $\{\EGFErbBfourxdim\}$,$\{\HRGErbBonexdim\}$,\\ \hspace*{1.4cm} $\{\HRGErbBtwoxdim\}$, $\{\HRGErbBthrxdim\}$, $\{\HRGErbBfourxdim\})$
\end{description}

\begin{table}[htb]
	\begin{center}
		{\small
			\caption{Two events modeling the forward and reverse directions of the third reaction of the $\ErbB$ signaling pathway}\label{tab:ErbB_model}
			\begin{tabular}{p{6cm}p{6cm}}
				\begin{tabular}{|l|} %|cllc|cllc|
					\hline
					\textbf{Rec3f}\\
					\textbf{WHERE}\\
					\ \ \textbf{@grd1} $\EGFEGFRdim \geq 1 $ \\
					\textbf{THEN} \\
					\ \ \textbf{@act1}\\
					\ \ \hspace{0.3cm}$\EGFEGFRdim :=\EGFEGFRdim - 1$\\
					\ \ \textbf{@act2}\\
					\ \ \hspace{0.3cm}$\EGFEGFRxdim :=\EGFEGFRxdim + 1$\\
					\textbf{END}\\
					\hline
				\end{tabular}
				
				&
				
				\begin{tabular}{|l|} %|cllc|cllc|
					\hline
					\textbf{Rec3r}\\
					\textbf{WHERE}\\
					\ \  \textbf{@grd1} $\EGFEGFRxdim \geq 1 $\\
					\textbf{THEN} \\
					\ \ \textbf{@act1}\\
					\ \ \hspace{0.3cm}$\EGFEGFRxdim := \EGFEGFRxdim - 1$\\
					\ \ \textbf{@act2}\\
					\ \ \hspace{0.3cm}$\EGFEGFRdim := \EGFEGFRdim + 1$\\
					\textbf{END}\\
					\hline
				\end{tabular}
				
			\end{tabular}
		}
	\end{center}
\end{table}

\vspace*{-1cm}
In the refined model, all events involving the variable-to-be-refined are replaced with new events. For example, consider the refinement of the events presented in the Table \ref{tab:ErbB_model}. 
%
%We define the relationship between the sets in the \emph{Context} part of the model and the corresponding variable in the \emph{Machine} part of the variable. Each variable present in the \emph{Machine} part will belong to its corresponding set present in the \emph{Context} part. This relationship makes sure that a variable can be replaced with each of the value set does have. For example, the relationship for sets $\EGFEGFRrdim$ and $\EGFEGFRxrdim$ defined in the \emph{Context} with variables $\EGFEGFRdim$ and $\EGFEGFRxdim$ defined in the \emph{Machine} part of the model is shown below:
%\begin{description}
%	\item {\bf @inv61}: $\EGFEGFRdim \in  \EGFEGFRrdim \longrightarrow \mathbb{N}$
%	\item {\bf @inv62}: $\EGFEGFRxdim \in  \EGFEGFRxrdim \longrightarrow \mathbb{N}$
%	\end{description}
The event \textbf{Rec3f} is replaced with event \textbf{Rec3f\_Ref} while the event \textbf{Rec3r} is replaced with event \textbf{Rec3r\_Ref}, shown in Table \ref{tab:ErbB_Refinedmodel}. 
%For event \textbf{Rec3f\_Ref}, the guard \textbf{grd1} shows that $\EGFEGFRdimh$ should be greater than or equal to one as it is get consumed by one (\textbf{act1}) and $\EGFEGFRxdimi$ is increased by one (\textbf{act2}). 
The dimer of $\EGFEGFRdim$ is refined to 8 different species. In this refinement strategy, we only consider homodimers (with their two components identical). The second guard \textbf{grd2} of these events is important for the refinement as it covers all the refinement scenarios. This guard has eight different conditions which are to cover all homodimers. The benefit of using functions is also shown in this event. Had we not used functions, we would have had to include 8 new events for the refinement of event \textbf{Rec3f} and similarly 8 new events for the refinement of event \textbf{Rec3f}. Without using functions the refined model of $\ErbB$ signalling pathway present in \cite{IANCU2019} would have $1320$ events but now it has $242$ events, as many as the basic model. It also has 53 axioms and 110 variables (same number of invariants as well). All proof obligations were discharged automatically.

\begin{table}[htb]
			\caption{Two events modeling the refinement of the forward and reverse directions of the third reaction of the $\ErbB$ signaling pathway}\label{tab:ErbB_Refinedmodel}	
{\scriptsize

			\begin{tabular}{p{5.9cm}p{6cm}}
				\begin{tabular}{|l|} %|cllc|cllc|
					\hline
					\textbf{Rec3f\_Ref}\\
					\textbf{ANY}\\
					h, i     \\
					\textbf{WHERE}\\
					\ \ \textbf{@grd1} $\EGFEGFRdimh \geq 1 $ \\
					\ \ \textbf{@grd2} \\
					\ \ \hspace*{0.2cm}$(h =  \EGFErbBonedim  \land i = \EGFErbBonexdim $) $\lor$ \\
					\ \ \hspace*{0.2cm}$(h =  \EGFErbBtwodim  \land i = \EGFErbBtwoxdim $) $\lor$ \\
					\ \ \hspace{0.2cm}$(h =  \EGFErbBthrdim  \land i = \EGFErbBthrxdim $) $\lor$ \\
					\ \ \hspace*{0.2cm}$(h =  \EGFErbBfourdim  \land i = \EGFErbBfourxdim $) $\lor$ \\
					\ \ \hspace{0.2cm}$(h =  \HRGErbBthrdim  \land i = \HRGErbBthrxdim $) $\lor$ \\
					\ \ \hspace*{0.2cm}$(h =  \HRGErbBfourdim  \land i = \HRGErbBfourxdim $) $\lor$ \\
					\ \ \hspace{0.2cm}$(h =  \HRGErbBthrdim  \land i = \HRGErbBthrxdim $) $\lor$ \\
					\ \ \hspace*{0.2cm}$(h =  \HRGErbBfourdim  \land i = \HRGErbBfourxdim $)  \\
					\textbf{THEN} \\
					\ \ \textbf{@act1} \\
					\ \ \hspace*{0.2cm}$\EGFEGFRdimh := \EGFEGFRdimh - 1$\\
					\ \ \textbf{@act2} \\
					\ \ \hspace*{0.2cm}$\EGFEGFRxdimi := \EGFEGFRxdimi + 1$\\
					\textbf{END}\\
					\hline
				\end{tabular}
			&
				\begin{tabular}{|l|} %|cllc|cllc|
					\hline
					\textbf{Rec3r\_Ref}\\
					\textbf{ANY}\\
					h, i     \\
					\textbf{WHERE}\\
					\ \  \textbf{@grd1} $\EGFEGFRxdimi \geq 1 $\\
					\ \ \textbf{@grd2} \\
					\ \ \hspace*{0.2cm}$(h =  \EGFErbBonedim  \land i = \EGFErbBonexdim $) $\lor$\\
					\ \ \hspace*{0.2cm}$(h =  \EGFErbBtwodim  \land i = \EGFErbBtwoxdim $) $\lor$ \\
					\ \ \hspace{0.2cm}$(h =  \EGFErbBthrdim  \land i = \EGFErbBthrxdim $) $\lor$ \\
					\ \ \hspace*{0.2cm}$(h =  \EGFErbBfourdim  \land i = \EGFErbBfourxdim $) $\lor$ \\
					\ \ \hspace{0.2cm}$(h =  \HRGErbBthrdim  \land i = \HRGErbBthrxdim $) $\lor$ \\
					\ \ \hspace*{0.2cm}$(h =  \HRGErbBfourdim  \land i = \HRGErbBfourxdim $) $\lor$ \\
					\ \ \hspace{0.2cm}$(h =  \HRGErbBthrdim  \land i = \HRGErbBthrxdim $) $\lor$ \\
					\ \ \hspace*{0.2cm}$(h =  \HRGErbBfourdim  \land i = \HRGErbBfourxdim $)  \\
					\textbf{THEN} \\
					\ \ \textbf{@act1} $\EGFEGFRxdimi := \EGFEGFRxdimi - 1$\\
					\ \ \textbf{@act2} $\EGFEGFRdimh := \EGFEGFRdimh + 1$\\
					\textbf{END}\\
					\hline
				\end{tabular}
			\end{tabular}
}
\end{table}

%It was not possible to build the full $\ErbB$ model using the approach in \cite{Usman_2017} as Rodin was not able to refine that big of a model. The use of functions has made it possible to build a model as big as the $\ErbB$ model in Rodin and to the best of our knowledge this is the biggest model ever built in Rodin. The number of events for the refined model of $\ErbB$ is 242 using functions. If Rodin has allowed us to build the full model of $\ErbB$ using the approach in \cite{Usman_2017}, the number of events in that case would have been 928 events in Rodin.

%\begin{itemize}
%\item Compare here the outcome of this approach with the outcome of the old approach. Compare number of variables and events. 	
%\end{itemize}

\section{Discussion}\label{sec:discussion}

%\begin{itemize}
%\item Summarise what we did
%\item Improvements in the scalability
%\item Limitations: Rodin did not allow us to implement the refinement of the ErBB model, it was too much with respect to the various glue invariants and proof obligations related to them.
%\item Further work: usability of these models, exploring their properties.
%\end{itemize}

Modeling and analyzing complex biological systems has never been as easy task. A solution to addressing complexity is to use refinement and start modeling from a conceptually simple  (abstract) model that is consistent: all its properties of interest hold. Then, we can gradually add all the necessary details in a correctness-by-construction approach, so that the most detailed (concrete) model still preserves all properties of interest. When models are large and complex, size becomes a bottleneck and it is simply unfeasible to model without tool support. Fortunately, Event-B is a (state-based) formal method built on the idea of refinement and has a suitable toolset - Rodin. However, when used liberally and without proper planning, even Rodin cannot handle arbitrarily large models. We have encountered this problem two years ago when trying to model the $\ErbB$ signaling pathway in Event-B and the concrete model had $1320$ reactions: Rodin could not handle it.

In this paper, we propose a modeling method that plays at Event-B's and Rodin's strengths: the high-level abstraction mechanisms, in particular using the common mathematical concept of function. The combinatorial explosion in the number of variables and events is generated in the case of data refinements. One species (protein, gene, etc) is to be replaced by a number of subspecies and each event involving the original species is refined by a set of events. This set is potentially as big as the number of subspecies or, if there is more than one species refined in one event, then the set can be as big as the product of the numbers of subspecies. Clearly, biology is so complex that we would have very soon a combinatorial explosion of variables modeling subspecies and events handling their reactions.

Our proposal is to go more abstract (`higher-level') and replace each species to be refined by a function defined on the constant set of all the subspecies. This simple artifact lets us express almost all the complexity in the event guards, where we can have many cases and combinations of parameters. Event-B and Rodin excel at handling guards and suddenly we have only a slight increase in the number of events, while the number of variables remains constant.

Our approach in this paper is demonstrated through two case studies, the heat shock response and the $\ErbB$ signaling pathway, both simplified enough to prove our point. Thus, we offer a proof-of-concept that our solution is viable, as its evaluation on the two case studies indicates.

The impact of this approach on future models of complex biological systems is significant. If we can start modeling from a conceptually simple but still consistent version of our system of interest, then we can add the necessary details via refinement, but so that we capture the main complexity in guards, via the high-level abstraction provided by functions. Size-wise, our models would remain manageable and still very expressive, albeit in a disciplined manner. Meanwhile, in this paper, we constructed the largest Event-B model ever built.

\paragraph{{\bf Acknowledgment}} Ion Petre was partially supported by the Romanian Ministry of Education and Research, CCCDI – UEFISCDI, project number PNIII-P2-2.1-PED-2019-2391, within PNCDI III.

\bibliographystyle{plain}
\bibliography{References}

\end{document}